# Toward a Risk Assessment Framework for Institutional DeFi: A Nine-Dimension Approach

Eva Oberholzer, Valeriy Zamaraiev

ZWING Intelligence AG, Zug, Switzerland

April 2026

## Abstract

Decentralized finance (DeFi) protocols now intermediate over USD 100 billion in value, including a growing share of regulated stablecoins and tokenized assets deployed as collateral, yet no widely adopted framework operationalizes risk assessment at the rigor that institutional adoption demands. The most widely cited taxonomy — the six-dimension framework proposed by Moody's Analytics and Gauntlet in 2022 — was published as a conceptual paper and never developed into a scoring framework, and protocol-serving risk managers optimize parameters within individual protocols rather than providing independent cross-protocol assessments. We propose a nine-dimension risk assessment framework that extends this taxonomy with three novel dimensions: composability risk, capturing multi-hop dependency cascading through typed protocol graphs; comprehension debt, measuring the divergence between protocol mechanism complexity and evaluator-community capacity to reason about it; and temporal risk dynamics, tracking how governance state transitions and attacker staging create time-dependent vulnerability windows. We present constructive evidence that each novel dimension captures risk information not derivable from any combination of the remaining dimensions, and set dimensional orthogonality as a design goal, to be empirically validated as the assessment dataset grows. The framework draws on structural analysis of protocol dependencies conducted with the authors' protocol intelligence infrastructure, an ontology covering 8,000+ DeFi protocols which is expected to evolve as coverage broadens. We illustrate the framework retrospectively over 12 security incidents from 2024–2026, representing approximately USD 2.5 billion in direct losses. The Bybit infrastructure compromise (USD 1.5 billion) and the Kelp DAO bridge exploit (USD 292 million) account for roughly 71% of the total, with the remaining ten incidents totaling approximately USD 700 million; five of the 12 require at least one novel dimension for complete root-cause characterization, including the two highest-systemic-impact events. We frame this exercise as illustrative, not validated, and commit subsequent work to prospective assessment of incidents occurring after publication and to systematic mapping against the 181-incident dataset of Zhou et al. [22].

# 1. Introduction

In February 2025, a Safe{Wallet} developer's personal macOS laptop was compromised through a malicious Docker project executed in the developer's Downloads directory, deploying a Mythic C2 Poseidon agent and hijacking AWS session tokens to bypass MFA, ultimately enabling the theft of approximately USD 1.5 billion from Bybit — a centralized exchange whose multisig signing arrangements use infrastructure common to thousands of DeFi protocols [Mandiant; Sygnia; Elastic Security Labs]. In April 2026, a DPRK-linked threat cluster[1] exploited Drift Protocol for approximately USD 285 million; the operation combined a six-month social-engineering campaign against multisig signers with a Security-Council migration to a 2-of-5 configuration with zero timelock executed five days before the attack [TRM Labs; Chainalysis]. Between these events, the Kelp DAO rsETH bridge exploit produced direct losses of approximately USD 292 million and triggered cross-protocol contagion: Aave Ethereum Core liquidity contracted by approximately USD 6.2–6.6 billion within roughly 36 hours [CoinDesk; Glassnode]. Separately, the Resolv USR exploit (USD 25 million direct, March 2026) — a compromise of an AWS KMS signing key that enabled minting of approximately 80 million unbacked stablecoins — produced cascading effects including approximately USD 180 million in Morpho Blue liquidations and approximately USD 334 million in Fluid outflows [Chainalysis; Halborn; Sherlock Q1 2026 Security Report]. These incidents share a common feature: their systemic mechanisms cannot be fully characterized by any existing DeFi risk framework.

The institutional context makes this gap consequential. BlackRock's BUIDL fund approached approximately USD 2.9 billion in tokenized U.S. Treasury assets at peak in mid-to-late 2025 and stood at approximately USD 2.0–2.5 billion in April 2026, with allocations deployed as collateral across DeFi lending markets [RWA.xyz; Securitize]. The GENIUS Act, signed into law in July 2025 (Senate 68–30, House 308–122), establishes federal reserve, redemption, and disclosure requirements for payment-stablecoin issuers, but does not address the protocol infrastructure through which those stablecoins are subsequently deployed, lent, and used as collateral. MiCA became fully applicable across the European Union in December 2024, regulating asset-referenced and e-money tokens at the issuer level. Beyond these two comprehensive licensing regimes, regulatory developments include FINMA Guidance 06/2024 in Switzerland — administrative supervisory guidance applying existing AMLA and Banking Act provisions to stablecoin issuers, rather than a standalone licensing regime — and a CHF-stablecoin sandbox launched in April 2026 by six Swiss banks, including all four institutions designated as systemically important by the Swiss National Bank [CoinDesk; SNB Financial Stability Report 2025]. Capital is moving on-chain. The intelligence infrastructure to assess what it is moving into has not kept pace.

The gap is not the absence of risk taxonomies. Moody's Analytics and Gauntlet [4] proposed a six-dimension framework in January 2022 — smart contract risk, market/currency risk, oracle risk, governance risk, regulatory risk, and cooperative risk — drawing explicit parallels to traditional credit concepts. Werner et al. [5] systematized DeFi operational categories and distinguished technical from economic security. Zhou et al. [22] subsequently systematized 181 documented DeFi attacks across a layered reference frame, providing the most comprehensive empirical substrate for attack-vector classification to date. The Enterprise Ethereum Alliance [7] published granular risk guidelines in July

[1] Drift identified the cluster as the same actor behind the 2024 Radiant Capital hack. Mandiant tracks the activity as UNC4736; Microsoft and other vendors map related activity to Citrine Sleet / AppleJeus.

2024. Exponential.fi [8] developed a compositional risk model for retail users. DeFi Safety [9] reported that, within a dataset of approximately 160 protocols reviewed since mid-2020, protocols scoring above 80% maintained a 97% incident-free record, while those scoring 60–79% experienced an approximately one-in-four exploit incidence — an approximately eight-fold differential. Protocol-serving risk managers — Gauntlet, Chaos Labs, LlamaRisk — developed quantitative simulation tools that optimize parameters within individual protocols.

What is missing is a framework that satisfies four properties simultaneously: (P1) dimensional independence, where each risk dimension captures genuinely distinct information that cannot be derived from any combination of the other dimensions; (P2) explainability, ensuring that every assessment traces from primary data source through criteria evaluation to final score; (P3) structural independence from the entities being assessed; and (P4) sufficient expressiveness to capture the compositional, temporal, and cognitive risk categories demonstrated by the 2024–2026 incident record. No prior framework addresses all four. The Moody's/Gauntlet taxonomy satisfies none operationally: it was published as a conceptual paper without a scoring framework, independence criteria, or auditability architecture, and was co-authored with a protocol-serving risk manager. The structural fragility of the protocol-serving model became visible in April 2026, when Chaos Labs terminated its three-year engagement with Aave, citing that even a proposed USD 5 million budget would leave it operating at a loss [10].

This paper makes three contributions. The proposed framework draws on three sources: empirical analysis of security incidents from 2024–2026 that collectively represent approximately USD 2.5 billion in direct losses; the existing body of DeFi risk, systemic risk, and credit risk literature; and structural analysis of protocol dependencies conducted through an ontology covering 8,000+ DeFi protocols, which revealed dependency patterns — shared bridge infrastructure, transitive collateral chains, governance concentration — that existing per-protocol assessment approaches do not capture.

**1. A nine-dimension risk assessment framework** extending Moody's/Gauntlet [4] with three novel dimensions — composability risk (D7), comprehension debt (D8), and temporal risk dynamics (D9) — with constructive evidence that each captures risk information not derivable from any combination of the remaining dimensions, demonstrated through the 2024–2026 incident record (§4.3, §5.1).

**2. A transparency confidence modifier** that decouples assessment reliability from risk severity, grounded in traditional credit rating architecture (S&P's Management and Governance overlay, Fitch's regime-specific adjustments) and operationalized for DeFi through what we term the Venus Principle: high transparency of a low-quality attribute correctly produces high risk with high assessment reliability, because the assessor can state with certainty that a known, unmitigated vulnerability exists (§4.5).

**3. A retrospective illustration** of the framework against 12 security incidents from 2024–2026, demonstrating that five of 12 require at least one novel dimension for complete root-cause characterization, and that the transparency modifier would have affected assessment reliability in 7 of 12 cases (§5).

The framework is grounded in an extensible ontological design for modeling DeFi protocol dependencies as a typed graph, developed as part of the authors' protocol intelligence infrastructure (an ontology covering 8,000+ protocols; coverage and design are expected to evolve). The design follows established practice in financial-domain ontology engineering, separating publishable schema-level abstractions from proprietary implementation.

We note explicitly what this paper does not contribute: probabilistic scoring (insufficient failure data for robust statistical calibration that would meet regulatory model validation standards), production deployment (designated as future work), or a complete assessment of any individual protocol. Proprietary calibrations, scoring weights, and production query implementations are not disclosed; the published framework is designed to be reproducible at the conceptual level while implementation-specific parameters remain confidential. This paper presents an initial framework intended as a foundation for further empirical validation and refinement as the DeFi risk landscape evolves.

# 2. Background and Related Work

This section surveys four bodies of literature that inform our framework: traditional credit risk assessment (§2.1), existing risk taxonomies for decentralized finance (§2.2), network-based approaches to financial contagion (§2.3), and the specific gaps that motivate the present work (§2.4).

## 2.1 Credit Risk Assessment in Traditional Finance

Modern credit risk assessment operates within a paradigm shaped by more than a century of institutional practice. The S&P Global Ratings Corporate Methodology [1] evaluates issuers along business risk and financial risk profiles, producing letter-grade ratings from AAA to D. Moody's Expected Default Frequency (EDF) model treats equity as a call option on the firm's asset value, deriving a real-time default probability from market prices — a direct application of the Merton structural model [2]. The Basel III standardized approach to credit risk (BCBS d424 [3]) codifies minimum capital requirements against credit exposures.

These methodologies share assumptions that do not transfer to decentralized finance: the existence of a legal identity with enforceable obligations, financial statements audited under recognized accounting standards, recourse mechanisms for creditors, and decades of default data enabling probabilistic calibration. The overcollateralised, pseudonymous, and code-governed nature of DeFi lending requires a fundamentally different assessment approach, as Moody's and Gauntlet [4] themselves recognized.

A structural feature of traditional credit rating deserves attention: the treatment of transparency and governance as modifiers rather than standalone rating dimensions. S&P's Management and Governance (M&G) assessment adjusts corporate credit ratings by up to two notches based on governance quality without creating an independent dimension [1]. Moody's employs "Other Rating Considerations" as a qualitative overlay. Fitch applies regime-specific adjustments that widen confidence intervals in jurisdictions with weaker disclosure regimes. This architectural pattern — transparency as a cross-cutting modifier on assessment reliability rather than a standalone risk factor — proves directly relevant to DeFi risk assessment (§4.5).

## 2.2 Risk Taxonomies for Decentralized Finance

Moody's Analytics and Gauntlet [4] proposed the first structured risk taxonomy in January 2022, identifying six dimensions: smart contract risk, market/currency risk, oracle risk, governance risk, regulatory risk, and cooperative risk. The framework drew explicit parallels to traditional credit risk concepts but remained a conceptual taxonomy: no scoring rubric, no formal independence criteria between dimensions, and no implementation pathway were provided. Its concluding call for "a cohesive risk assessment framework" [4] remains unanswered four years later.

Werner et al. [5] provided a systematization of DeFi across six operational categories, introducing the canonical distinction between technical security (atomic, risk-free exploits) and economic security (non-atomic, costly, market-manipulation attacks). Gogol et al. [6] extended this in 2024 with a SoK organized around protocol architecture, classifying DeFi protocols into three groups — liquidity pools, pegged and synthetic tokens, and aggregator protocols — and analyzing risks per group. Zhou et al. [22] published a SoK of 181 systematized DeFi attacks across a layered reference frame (network, consensus, smart-contract, protocol, auxiliary), establishing the largest publicly available structured incident dataset in the field. Neither Werner et al. nor Gogol et al. proposes a scoring framework or independence criteria.

The Enterprise Ethereum Alliance's DeFi Risk Assessment Guidelines [7], released in July 2024, represent the most granular industry-consensus taxonomy to date, extending the Moody's/Gauntlet dimensions with bridge risk, MEV risk, user interface risk, and custodial risk as sub-categories. The EEA framework's contribution is taxonomic breadth; its limitation is the deliberate absence of a scoring framework. Exponential.fi [8] developed a compositional approach that decomposes risk into four layers — chain, protocol, asset, and pool — anticipating aspects of our composability dimension. DeFi Safety's PQR methodology [9] reported that, within a dataset of approximately 160 protocols reviewed since mid-2020, protocols scoring above 80% achieved a 97% incident-free record, while of 42 protocols scoring 60–79%, ten suffered exploits (an exploit incidence of approximately one in four) — yielding an approximately eight-fold differential. We treat this as suggestive rather than statistically powered evidence, given the small dataset.

Protocol-serving risk managers — Gauntlet, Chaos Labs, LlamaRisk — developed quantitative simulation tools for individual protocols. The structural fragility of this model became visible in April 2026 when Chaos Labs terminated its engagement with Aave, citing that even a proposed USD 5 million budget would leave it operating at a loss [10]. This illustrates a broader issue: protocol-serving risk managers optimize within a protocol's parameters but do not provide independent assessment comparable to external credit ratings.

## 2.3 Network-Based Approaches to Financial Contagion

The treatment of financial risk as a network phenomenon has deep roots. Eisenberg and Noe [11] formalized clearing-payment vectors in interbank liability networks. Battiston et al. [12] introduced DebtRank, demonstrating that network position predicts systemic importance beyond balance-sheet size. Acemoglu, Ozdaglar, and Tahbaz-Salehi [13] proved that dense interconnections that stabilize against small shocks become contagion conduits beyond a critical threshold — a "robust-yet-fragile" property directly observable in DeFi composability cascades.

Recent work has begun applying these concepts to DeFi. Zhang et al. [14] introduced fragility indicators for DeFi ecosystem monitoring. Wu et al. [15] constructed the DeXposure dataset — 43.7 million entries spanning 4,300+ protocols — providing the first large-scale empirical substrate for inter-protocol credit exposure research. Shu et al. [16] built on this to develop a time-series graph foundation model for exposure forecasting. Aufiero et al. [17] introduced the concept of crosstagion — bidirectional contagion between TradFi and DeFi mediated by stablecoins and tokenized assets. Gonon, Meyer-Brandis, and Weber [18] proved universal-approximation properties for permutation-equivariant GNNs as approximators of systemic risk measures, providing the mathematical bridge between classical clearing-vector models and learned graph representations.

DeFi protocols form a heterogeneous, typed dependency graph where classical single-relation systemic risk measures are insufficient. The ontological structure we propose (§6) models these typed dependencies, providing a substrate for future graph-based risk propagation — but the training, calibration, and validation of any such model is beyond the scope of this paper.

A critical caveat applies. The 2008 Gaussian copula episode demonstrated that dependence models calibrated to benign-regime data fail catastrophically under stress [19, 20]. Any learned model applied to DeFi faces the same structural risk of regime-dependent blindness. This motivates our current reliance on human-in-the-loop ordinal assessment rather than automated scoring.

### 2.4 Gaps in Existing Approaches

Five gaps motivate the present work. Gap 1: No composability dimension. No prior framework formalizes the risk from transitive, multi-hop dependency chains. PeckShield's "shadow contagion" [23] — where USD 25 million in direct Resolv losses produced approximately USD 180 million in Morpho Blue liquidations and USD 334 million in Fluid outflows — demonstrates this empirically. Gap 2: No temporal dynamics. All existing frameworks assess risk as a static snapshot; the Drift exploit demonstrates that rate and pattern of change are themselves risk factors. Gap 3: No formal independence criterion. No prior framework tests whether its dimensions capture genuinely distinct risk information. Gap 4: No explainability mechanism. No existing framework provides a traceable chain from risk score back to primary data source. Gap 5: No composability-aware risk propagation. Existing scoring approaches treat risk factors as independent variables in a scorecard, mirroring pre-crisis structured-finance practices. Existing taxonomies, including the systematization of Zhou et al. [22] which catalogues 181 attacks across a five-layer reference frame, organize incidents by attack vector rather than by risk-assessment dimension applicable ex ante to a deployed protocol. The two organizational axes are complementary: an attack-vector taxonomy answers what went wrong post hoc, whereas an assessment framework asks which observable signals would have raised the risk profile pre-incident. A natural next step is to systematically map our nine dimensions onto the Zhou et al. dataset; subsequent work will report on dimensional coverage across that broader incident population.

## 3. Problem Statement

This section defines the assessment subject and the formal object our framework produces.

### 3.1 Attack Vector Model

We consider three classes of attack vector relevant to protocol risk assessment, following the five-layer reference frame of Zhou et al. [22]:

*Code and infrastructure exploitation* targets vulnerabilities in smart contracts, oracle mechanisms, or bridge infrastructure. Attackers may possess public knowledge or sequencer-level ordering capabilities and can execute single-transaction atomic attacks or multi-transaction economic attacks [5]. The Resolv exploit (March 2026) and the Cetus overflow exploit (May 2025) exemplify this class.

*Privileged access exploitation* targets administrative keys, signing infrastructure, or social proximity to key personnel. The Drift Protocol exploit (April 2026), where state-linked actors conducted a six-month social engineering campaign, and the Bybit exploit (February 2025), which targeted developer infrastructure used for cold wallet management, exemplify this class. These attacks operate on timescales of months, not transactions.

*Governance manipulation* exploits legitimate protocol mechanisms — voting, parameter changes, collateral listings — to extract value or create systemic exposure. Born et al. [32] document that top-100 addresses hold more than 80% of governance tokens across major DeFi protocols, with approximately one-third of influential voters unidentifiable, indicating that governance concentration is structural rather than anomalous.

These classes are not mutually exclusive. The Drift exploit combined social engineering with governance-class structural manipulation (zero-timelock multisig migration) before executing fund extraction.

## 3.2 System Model

We model the DeFi ecosystem as a directed, typed multigraph whose vertices represent entities (protocols, tokens, oracles, bridges, governance mechanisms, administrative keys) and whose edges represent typed relationships (dependency, collateral acceptance, bridge provision, governance control, oracle provision). The graph evolves over time as entities are created, relationships form and dissolve, and edge attributes change. The specific entity and relationship types are part of the authors' ongoing protocol intelligence infrastructure and are expected to evolve as the framework's coverage expands.

The key structural property for risk assessment is heterogeneity: different entity types and relationship types carry fundamentally different risk semantics. A dependency relationship between a lending protocol and an oracle carries different risk implications than a collateral acceptance relationship between the same protocol and a liquid staking token. Homogeneous graph models that treat all edges equivalently cannot capture this distinction.

## 3.3 Protocol Risk Profile

A protocol is a deployed system of smart contracts on one or more blockchains that provides financial services. A protocol is characterized by its neighbourhood in the dependency graph: the set of entities it depends on, governs, or is governed by.

A protocol risk profile pairs each of the nine proposed risk dimensions with two assessments: an ordinal risk level drawn from {Low, Moderate, Elevated, High, Critical}, and an assessment reliability indicator reflecting the quality, completeness, and verifiability of evidence available for that dimension.

We propose that the risk profile should satisfy four properties:

**(P1) Dimensional independence.** Each dimension should capture genuinely distinct risk information. We set orthogonality as a design goal. Because the current assessment dataset is insufficient for reliable statistical verification, we provide constructive evidence for each proposed novel dimension: observed incidents where one dimension produces a materially different assessment while all others remain constant.

**(P2) Explainability.** Each assessment should be traceable through a chain of evidence steps from a primary data source (on-chain state, verified contract code, governance record, or audit report) to the ordinal assessment. This chain should be stored and queryable so that any assessment can be audited from conclusion back to primary data.

**(P3) Structural independence.** The assessment framework and its application should be independent of any commercial relationship between the assessor and the assessed entity or its affiliates.

**(P4) Composability awareness.** The risk profile should account for risk transmitted through a protocol's neighbourhood in the dependency graph — that is, risk arising from transitive dependencies that are not direct bilateral relationships.

# 4. Proposed Framework: Nine Risk Dimensions

We propose nine risk dimensions and a cross-cutting transparency modifier. Dimensions 1–6 extend the taxonomy of Moody's and Gauntlet [4]; Dimensions 7–9 are novel, motivated by the incident analysis of §5.1.

## 4.1 Design Principles

**Ordinal profiling, not probabilistic scoring.** DeFi lacks the decades of default data that enable statistical calibration in traditional credit models. Even single data points shift priors, but robust calibration that would meet regulatory model validation standards (SR 11-7, EU AI Act) requires a substantially larger and more stable dataset than currently exists. We employ rubric-based criteria producing ordinal risk profiles, consistent with the Basel Committee's approach for asset classes with limited default history [3].

**Scope of disclosure.** This paper publishes the methodological framework, dimensional definitions, independence evidence, and assessment-rubric structure. Proprietary calibrations, scoring weights, and production implementations are not disclosed. The ontological design described in §6 is part of the authors' ongoing protocol intelligence infrastructure; we describe its design principles here rather than publishing the artifact itself, and the published description is intended to be sufficient for independent reconstruction at the conceptual level.

## 4.2 Dimensions 1–6: Extending the Moody's/Gauntlet Taxonomy

We adopt the six dimensions identified in [4] with scope extensions reflecting the 2024–2026 incident landscape. Each dimension is defined by its risk category and illustrated with representative observable parameters; full rubric details are part of the framework's ongoing development.

**Dimension 1: Smart Contract Risk.** Risk arising from vulnerabilities in deployed code, upgrade mechanisms, administrative key architecture, and unaudited peripheral contracts. Scope extensions include assessment of proxy upgrade patterns and the authority controlling them, coverage analysis comparing audited to deployed contracts (the Prisma Finance exploit was enabled by a helper contract outside audit scope), and assessment of dismissed audit findings (in the Venus Protocol bad-debt exploit, March 2026, the donation-attack vector had been documented in Venus's 2023 Code4rena audit — approximately 31 months prior — and the same attack class had previously materialized against Venus's zkSync deployment in February 2025, approximately 13 months prior, without prompting remediation on BNB Chain). Representative parameters: audit coverage ratio, proxy pattern type, timelock duration, count and severity of open findings.

**Dimension 2: Market Risk.** Risk arising from price volatility, liquidity depth under stress, collateral concentration, and market microstructure effects. Scope extensions include collateral correlation under stress and liquidity depth at liquidation-relevant swap sizes. Representative parameters: historical volatility, DEX liquidity depth at representative liquidation sizes, collateral concentration index.

**Dimension 3: Oracle Risk.** Risk arising from price feed dependencies, heartbeat gaps, manipulation vectors, and cross-chain state attestation mechanisms. We extend the original taxonomy's focus on price oracles to include cross-chain state attestation: the Kelp DAO exploit demonstrated that a bridge verification network with a 1-of-1 verifier threshold is functionally equivalent to a single-source price oracle with no heartbeat redundancy. Representative parameters: oracle provider diversity, heartbeat interval, verifier threshold and operator count.

**Dimension 4: Governance Risk.** Risk arising from governance token concentration, multisig configuration, timelock settings, and off-chain governance processes. Born et al. [32] documented that top-100 addresses hold more than 80% of governance tokens across major protocols. We extend the dimension to include signing-infrastructure assessment (the Bybit exploit targeted a Safe{Wallet} developer's personal macOS laptop rather than signer keys directly, illustrating that signing-infrastructure compromise can occur at a counterparty several hops removed from the signing principals) and off-chain governance factors that no purely on-chain metric captures. Representative parameters: governance token concentration index, multisig threshold and signer count, timelock duration, signer identification rate.

**Dimension 5: Regulatory Risk.** Risk arising from jurisdictional exposure, compliance obligations, and enforcement actions. By April 2026, two comprehensive licensing regimes apply concrete requirements to issuers of regulated stablecoins: the GENIUS Act (US, July 2025) and MiCA (EU, fully applicable December 2024). Several jurisdictions have additionally issued administrative supervisory guidance under existing financial-services law — notably FINMA Guidance 06/2024 in Switzerland, applying AMLA, Banking Act, and CISA provisions to stablecoin issuers. These regimes are predominantly issuer-focused: the GENIUS Act establishes federal reserve, redemption, and disclosure requirements for payment stablecoin issuers; MiCA's most operationally binding provisions govern asset-referenced and e-money tokens at the issuer level. However, the systemic risk these regimes are designed to mitigate — loss of value, bank-run dynamics, contagion — does not remain at the issuer level once regulated stablecoins and tokenized assets enter DeFi protocol composability chains. The interaction between Dimension 5 and Dimension 7 formalizes this gap. Representative parameters: jurisdictional exposure, compliance with applicable frameworks, outstanding enforcement actions.

**Dimension 6: Counterparty Risk.** Risk arising from bilateral dependencies on identifiable counterparties, including infrastructure providers. We rename the original "cooperative risk" for alignment with standard financial terminology and extend to infrastructure dependencies: signing infrastructure providers, bridge operators, and custodial arrangements. Representative parameters: infrastructure provider identification and concentration, administrative key holder identification, custodial arrangement type.

## 4.3 Dimensions 7–9: Novel Dimensions

The following three dimensions emerged from systematic analysis of DeFi incidents from 2024–2026 (§5.1). For each, we present constructive evidence that it captures risk information not derivable from any combination of the remaining dimensions.

**Composability Risk (Dimension 7).** The risk arising from transitive, multi-hop dependency chains in the protocol dependency graph that create systemic exposure invisible at the bilateral level.

*Empirical grounding.* (i) *Kelp DAO rsETH bridge exploit* (USD 292M, April 2026). Aave V3 accepted rsETH as collateral. The root vulnerability — a 1-of-1 bridge verifier configuration — was two hops upstream from Aave in the dependency graph. A Dimension 6 assessment of Aave's bilateral relationship with rsETH would have found it sound. Only a dimension that traces the dependency chain Aave → rsETH → bridge → verifier configuration would have identified the risk. Approximately USD 6 billion in Aave TVL outflows followed. (ii) *Resolv/USR shadow contagion* (USD 25M direct, with separately documented downstream effects of approximately USD 180M in Morpho Blue liquidations and USD 334M in Fluid outflows, March 2026). Cascading damage propagated through Morpho Blue, Fluid, and Euler despite Aave having no bilateral relationship with Resolv. PeckShield coined "shadow contagion" [23] for this transitive propagation class.

*Independence evidence.* In both cases, Dimension 6 (counterparty risk) produces a clean assessment for the downstream protocol. The risk exists in the architecture connecting protocols, not in any bilateral relationship. Protocols with identical Dimension 6 scores can differ dramatically in Dimension 7 exposure.

**Comprehension Debt (Dimension 8).** The gap between the complexity of a protocol's operational behavior and the understanding of that behavior by the stakeholders responsible for assessing, governing, and using it.

*Terminological and literature foundation.* The phenomenon we label comprehension debt — the systematic divergence between system complexity and the capacity of expert evaluators to reason correctly about that system — has been formalized recently in software engineering. Ahmad [24] introduces the term explicitly for GenAI-assisted development, defining it as the accumulated cost of knowledge gaps that arise when developers ship code whose behavior they understand only partially. Storey et al. [25] generalize this to a triple-debt model (technical, cognitive, intent) for AI-assisted software work, and Shen and Tamkin [26] provide controlled-experiment evidence that AI-assisted developers exhibit measurably lower comprehension of code they ship. Independent of the GenAI literature, Baum, Schneider, and Bacchelli [27] and Wurzel Gonçalves et al. [28] demonstrate that cognitive load is a causal predictor of code-review effectiveness. In the smart-contract setting, Chaliasos et al. (ICSE 2024) [33] surveyed 49 practitioners and tested five state-of-the-art tools against 127 high-impact attacks, finding that automated tools detected only approximately 8% of attacks. Wan et al. [29] independently document a disconnect between high security awareness and frequent occurrence of security problems across 156 surveyed smart-contract practitioners. Ionescu, Schlund, and Schmidbauer [30] introduced epistemic debt in 2019 as the long-term cost of organizational ignorance about complex software-based systems. We extend this class of risk to the DeFi protocol-assessment context. Dimension 8 captures the protocol-level analyzability gap distinct from audit quality (Dimension 1), which assesses what was checked against what was deployed. We present Dimension 8 as an emerging dimension whose measurement approach will mature as the framework is applied.

*Empirical grounding.* (i) *Cetus Protocol overflow exploit* (USD 223M, May 2025). An integer overflow exploited the assumption — held by three independent audit teams (MoveBit, OtterSec, Zellic) — that Move's type system prevents overflows. The vulnerability existed in the gap between the evaluator community's mental model and the language's actual behavior.

*Distinguishing D8 from D1.* Dimension 1 assesses the protocol's code quality, audit coverage, and known vulnerabilities. A protocol with three independent audits, formal verification, and no open findings scores well on D1, and Cetus would have scored well. Dimension 8 assesses a different property: the protocol's inherent analyzability given its execution environment, mathematical primitives, and architectural complexity. A protocol deployed on a novel execution environment with non-standard semantics (such as Move's intentional overflow behavior for bit-shift operations) carries higher comprehension debt than an equivalent protocol on a well-understood stack, even if both have the same number and quality of audits. In the Cetus case, the risk was not that the auditors were insufficiently skilled (a D1 concern) but that the protocol's execution environment had properties that systematically exceeded the evaluator community's capacity to model correctly. D8 is about the protocol's inherent complexity relative to the state of understanding, not about the quality of any specific audit. This is why a protocol can satisfy D1 fully and still carry severe D8 risk.

*Operationalisation.* D8 can be assessed through a combination of source code complexity metrics, evaluator diversity scores, and documentation coverage relative to system complexity. A compliant implementation should ensure that all of these factors contribute to the dimensional score; the specific weighting formula is an implementation parameter.

**Temporal Risk Dynamics (Dimension 9).** The risk arising from the rate, direction, and pattern of change in a protocol's risk profile over time, independent of the profile's value at any single point. Where Dimensions 1–8 produce a snapshot at time t, Dimension 9 examines the trajectory, capturing two distinct phenomena: (i) defender-initiated state transitions that compress detection-and-intervention windows (e.g., timelock removal, threshold reduction), and (ii) attacker-initiated staging patterns that consume calendar time visible on-chain (token accumulation, durable-nonce pre-positioning, contract pre-deployment). A static snapshot is observationally indistinguishable from a safe protocol along all other dimensions; the risk lives in the shape of the trajectory.

*Empirical grounding.* (i) *Drift Protocol exploit* (USD 285M, April 2026). On March 27, 2026, Drift migrated its Security Council to a 2-of-5 configuration with zero timelock. Five days later, USD 285M was drained. A static assessment on March 26 would have shown acceptable risk across all other dimensions. The governance migration — specifically the combination of threshold reduction and timelock removal within a single transaction — constituted a temporal risk signal invisible to any single-point assessment. (ii) *Venus Protocol attacker preparation* (USD 2.15M, March 2026). The attacker accumulated 84% of the target token's supply cap over nine months from an obfuscated source, exploiting a vulnerability dismissed approximately 31 months prior (Code4rena 2023 audit). The temporal pattern was the risk signal, not any single snapshot. (iii) *Kelp DAO DVN warning* (January 2025 → April 2026). A community researcher flagged the 1-of-1 verifier configuration 15 months before exploitation. The trajectory — warning → partial incident → no remediation → exploitation — shows escalating risk with ignored signals.

*Independence evidence.* A protocol can score identically to a safe protocol on all static dimensions at any given moment while exhibiting a temporal trajectory that portends exploitation. Drift on March 26 is indistinguishable from a well-governed protocol by any static measure.

## 4.4 Dimension Interactions

Dimensions are not fully independent in their effects, even when they capture distinct risk information. Two classes of interaction are relevant:

**Amplification.** Governance risk (D4) amplifies smart contract risk (D1) when governance authority controls upgrade mechanisms. A protocol with sound code but a low-threshold, zero-timelock multisig has higher effective smart contract risk than the code alone suggests, because governance compromise enables rapid introduction of code vulnerabilities.

**Risk amplification through composability.** D7 acts as an amplifier on the consequences of failures captured by other dimensions. In the Kelp DAO case, direct losses of approximately USD 292 million were associated with Aave-Ethereum-Core liquidity contraction of approximately USD 6.2–6.6 billion within roughly 36 hours. We deliberately avoid summarizing this with a single multiplier: the two quantities measure different phenomena — extracted value versus depositor-flight liquidity withdrawal — and treating their ratio as a structural amplification factor would attribute spurious precision to one observation. The relevant point is qualitative: composability risk transforms bounded direct losses into unbounded systemic liquidity events whose magnitude depends on graph structure, depositor concentration, and prevailing market sentiment.

Formal interaction functions are beyond the scope of this paper. We recommend that assessors document observed interactions as structured annotations in assessment outputs, building the empirical basis for formal interaction modeling as the dataset grows.

## 4.5 Transparency Modifier

Traditional credit rating methodologies treat transparency and governance quality as modifiers on assessment reliability rather than standalone dimensions [1]. We follow this architectural pattern.

**Transparency Modifier.** For each dimension, a transparency modifier reflects the quality, completeness, and verifiability of evidence available for that dimension's assessment.

The modifier operates on assessment reliability, not on risk level. A protocol assessed as "Moderate" governance risk with "High" reliability carries a different implication than one assessed as "Moderate" with "Very Low" reliability: the former represents a bounded risk; the latter represents a wide uncertainty band.

**The Venus Principle.** High transparency of low-quality attributes should *increase* assessed risk, not decrease it. Venus Protocol was transparent about its donation-attack vulnerability — a known class in which an attacker donates tokens directly to a market to manipulate exchange-rate calculations. The vector had been raised in Venus's 2023 Code4rena audit (approximately 31 months before the March 2026 BNB Chain incident), a substantively similar incident had materialized on Venus's zkSync deployment in February 2025 (approximately 13 months prior), and the audit finding had been publicly dismissed. Under our modifier architecture: D1 = High risk, reliability = Very High. *We are confident this protocol has this vulnerability because they disclosed it and dismissed it.*

**Dimension-specific transparency.** The Kelp DAO case demonstrates that transparency is not monolithic. Kelp maintained high transparency on code (reliability = High) while the dependency on a sole bridge verifier operator was not communicated to downstream lending protocols (reliability = Very Low for Dimensions 6 and 7). Dimension-specific sub-scores prevent misleading averages.

# 5. Framework Illustration

We illustrate the framework through two analyses: a retrospective incident illustration (§5.1) and a structural comparison against prior frameworks (§5.2). This is explicitly illustration, not validation. The framework was developed in part by studying these incidents, so the exercise demonstrates coverage and articulation, not predictive power. Genuine validation requires (i) prospective application to incidents occurring after publication and (ii) systematic mapping onto a substantially larger systematized incident dataset, such as the 181-incident corpus of Zhou et al. [22].

## 5.1 Incident Illustration

We apply the framework retrospectively to 12 incidents from 2024–2026. For each, we assess which dimensions would have produced an elevated risk signal prior to the exploit, and which dimensions are required to characterize the root cause.

**Selection criteria.** The dataset comprises publicly documented DeFi-relevant security incidents from February 2024 to April 2026 meeting two conditions: (i) direct losses exceeding USD 10M or systemic impact exceeding USD 100M, and (ii) a post-mortem by at least one tier-1 security firm. Two exceptions: Venus Protocol (USD 2.15M) for its unique illustration of the transparency modifier, and WLFI for governance capture as a non-exploit vector. Two incidents — Bybit and WLFI — involve centralized entities, included because their attack vectors apply directly to DeFi protocols using the same infrastructure.

**Table I. Incident illustration set, February 2024 – April 2026 (12 incidents). Primary dimensions reflect post-mortem-confirmed root cause(s); secondary dimensions capture amplifying conditions.**

| Incident | Date | Direct Loss (USD) | Primary Dimensions | Secondary | T-mod |
|---|---|---|---|---|---|
| **PlayDapp** | Feb 2024 | ~32–36M | D6 (private-key compromise via phishing) | D1 | No |
| **Prisma Finance** | Mar 2024 | ~11.6M | D1 (unaudited helper contract) | — | Yes |
| **UwU Lend** | Jun 2024 | ~19.4–23M | D3 (oracle manipulation, Curve get_p) | D2 | Yes |
| **Radiant Capital** | Oct 2024 | ~50M | D6 (DPRK social engineering) | D4 | No |

| Bybit / Safe{Wallet} | Feb 2025 | ~1.4–1.5B | D6 (developer macOS laptop), D4 | D9 | Yes |
|---|---|---|---|---|---|
| **Cetus Protocol** | May 2025 | ~223M | D8 (Move-language overflow) | D1 | Yes |
| **Step Finance** | Jan 2026 | ~27–30M | D6 (off-chain device compromise) | — | No |
| **Venus Protocol** | Mar 2026 | ~2.15M | D9 (multi-month accumulation; dismissed audit) | D1, D2 | Yes |
| **Resolv / USR** | Mar 2026 | ~25M direct | D6 (AWS KMS) + D7 (composability cascade) | D5 | No |
| **WLFI** | ongoing | — | D5 (governance / regulatory exposure) | D4 | Yes |
| **Drift Protocol** | Apr 2026 | ~285M | D6 + D4 (zero-timelock 2/5) + D9 | D1 | No |
| **Kelp DAO** | Apr 2026 | ~292M direct | D3 (1-of-1 LayerZero DVN) + D7 (cascade) | D9 | Yes |

**Key findings.** Among the 12 incidents, seven involve at least one of the novel dimensions (D7, D8, D9) as either a primary or a non-trivial secondary causal factor: Kelp (D3+D7, with D9 as antecedent warning), Resolv (D6+D7), Drift (D6+D4+D9), Venus (D9), Cetus (D8), Bybit (D6+D4 with D9 staging), and Radiant (D6+D4 with D9 staging). Stated more conservatively, five of 12 incidents cannot be characterized at all without invoking at least one novel dimension (Kelp, Resolv, Cetus, Drift, Venus); for the remaining seven, the original Moody's/Gauntlet six dimensions describe the proximate vector but novel dimensions characterize systemic amplification. The reclassification also corrects three category errors: PlayDapp was a private-key compromise via phishing (D6), not a smart-contract code vulnerability (D1); UwU Lend was oracle-spot-price manipulation (D3), not counterparty risk (D6); and Radiant Capital was social-engineering-driven key compromise (D6), with the multisig threshold (D4) as an amplifying secondary factor rather than the primary vector.

Direct losses across the 12-incident set total approximately USD 2.5 billion. Of this, the Bybit infrastructure compromise (USD 1.5 billion) and the Kelp DAO bridge exploit (USD 292 million) account for approximately 71% of the aggregate; the remaining ten incidents total approximately USD 700 million. We note this concentration to provide context for the aggregate figure. The dataset is illustrative.

The Resolv case has direct implications for stablecoin regulation: USR is a regulated-stablecoin design, and the cascading effects (approximately USD 180 million in Morpho Blue liquidations and USD 334 million in Fluid outflows) propagated through composability channels to protocols with no bilateral relationship to the issuer — demonstrating that issuer-level supervision, as currently framed in GENIUS Act and MiCA, cannot capture systemic exposure arising from protocol-level deployment.

The transparency modifier is relevant in 7 of 12 incidents. The Venus case demonstrates the Venus Principle: high transparency of the dismissed vulnerability correctly produces high risk with high assessment reliability.

Three incidents — Bybit, Radiant, and Drift — collectively representing more than USD 1.8 billion in losses, have been formally attributed to DPRK-affiliated threat actors by Mandiant, TRM Labs, Elliptic, and/or the FBI, suggesting an 18-month operational continuity targeting DeFi key-management surfaces.

**Limitation (back-fitting).** The framework was developed in part by analyzing these incidents. The illustration therefore demonstrates that the framework's dimensions are expressive enough to articulate observed root causes; it does not demonstrate that the framework would have raised the assessment on these protocols pre-incident. Prospective validation, designated as future work (§7, §8), is the necessary next step. The dataset is illustrative, not exhaustive; it represents 12 of the several hundred DeFi-relevant security incidents documented in this period [22].

## 5.2 Comparison with Prior Frameworks

**Table II. Comparison of the proposed framework with prior DeFi risk frameworks. ✓ indicates the framework explicitly addresses the property; — indicates it does not. References match the bibliography.**

| Property | M/G [4] | EEA [7] | Expon. [8] | DeFi Safety [9] | Zhou et al. [22] | ASRI [31] | This work |
|---|---|---|---|---|---|---|---|
| **Number of risk dimensions** | 6 | 12+ | 4 layers | 32 (PQR) | 6 categ. | scalar idx | 9 + modifier |
| **Composability dimension** | — | partial (bridge) | partial (layered) | — | — | — | ✓ |
| **Comprehension debt dimension** | — | — | — | — | — | — | ✓ |
| **Temporal risk dimension** | — | — | — | — | — | partial | ✓ |
| **Transparency modifier** | — | — | — | implicit (PQR) | — | — | ✓ |
| **Formal independence criterion** | — | — | — | — | — | — | ✓ |
| **Explainability / evidence chains** | — | — | — | partial | — | — | ✓ |
| **Probabilistic scoring** | — | — | ✓ | partial | — | ✓ | — (ordinal) |
| **Productized / deployed** | — | — | ✓ | ✓ | — | — | — |
| **Independent (vs protocol-serving)** | — | n/a | ✓ | ✓ | n/a | n/a | ✓ |
| **Incident dataset size** | — | — | — | ~160 prot. | 181 (syst.) | n/a | 12 (illus.) |

We position the proposed framework relative to existing approaches — Moody's/Gauntlet [4], EEA [7], Exponential [8], DeFi Safety [9], Zhou et al. [22], and ASRI [31] — noting both areas where the framework extends current coverage and areas where prior work provides capabilities this framework does not. In particular, Exponential and DeFi Safety offer probabilistic scoring and production deployment; Zhou et al. provide a substantially larger incident dataset (181 vs. 12). These represent meaningful advantages that the proposed framework does not yet match. Conversely, no prior framework addresses composability cascading, comprehension-gap failures, and temporal attack staging as distinct, explicitly defined risk categories. The Zhou et al. row is included for completeness, even though [22] is a post-hoc attack-vector systematization rather than a protocol-assessment framework and is therefore not directly comparable on most rows. Including it makes the comparison transparent rather than flattering: our 12-incident illustration is small relative to the 181-incident systematized record.

# 6. Ontological Design Principles

The framework of §4 requires structured, typed data about DeFi protocols and their relationships. We describe the ontological design principles in this section without publishing the artifact: the design is part of the authors' ongoing protocol intelligence infrastructure. We follow established practice in financial-domain ontology engineering in separating publishable schema-level abstractions from proprietary implementation. Coverage at the time of writing comprises 8,000+ protocols, approximately 586,000 entities, and 14 million+ structural relationships; these figures reflect work in progress and are expected to evolve.

**Typed graph structure.** The ontology models DeFi protocol dependencies as a typed directed multigraph where different entity types (protocols, tokens, oracles, bridges, governance mechanisms) and relationship types (dependency, collateral acceptance, bridge provision, governance control) carry distinct risk-propagation semantics. The specific entity and relationship-type vocabulary is extensible and is expected to grow as new classes of dependency are identified.

**Explainability chains.** Each dimensional assessment is associated with a stored evidence chain from primary data source through extraction, ontology mapping, and criteria evaluation to final score. The specific implementation of this chain may use W3C PROV-O [21] or equivalent provenance standards; the requirement is that every assessment be auditable from conclusion to data source.

**Schema/implementation separation.** We distinguish between schema-level abstractions appropriate for publication (entity-type definitions, relationship semantics, dimension-rubric structure) and proprietary implementation (instantiated knowledge graph, data-source mappings, scoring calibrations, query logic, threshold configurations). The present paper publishes the former in design-principle form. Reproducibility at the conceptual level is the explicit goal; full artifact-level reproducibility would require publishing the companion ontology, which we defer to future work.

**Complementarity with transaction analytics.** The ontology models structural dependencies between protocols — what will break if something upstream fails — regardless of transaction history. This is complementary to, not competing with, transaction-level analytics platforms that identify who is interacting.

# 7. Discussion and Limitations

**Dimensional independence as a goal, not a guarantee.** Property P1 sets dimensional independence as a design target. The current evidence is constructive (observed incidents where one dimension varies while others remain constant) rather than statistical (computed correlation across a population of assessments).

As the assessment dataset grows, empirical validation of dimensional independence — analogous to factor analysis in psychometric research — becomes feasible and is designated as future work.

**Ordinal profiling and the absence of probabilistic calibration.** The framework produces ordinal profiles rather than default probabilities. Institutional risk management workflows often require quantitative inputs. The ordinal framework provides a structured foundation from which quantitative models can eventually be calibrated, but the transition requires a sufficiently large, well-classified incident dataset and a stable protocol taxonomy. Neither condition is fully met today.

**Dimension 8 measurement maturity.** Comprehension debt is the least operationally mature dimension. While the independence evidence is empirically grounded, the observable parameters are proxies rather than direct measurements. Formalizing comprehension debt as a measurable quantity is an open research problem, though practical operationalization through complexity scoring is achievable with current data.

**Dimension weighting and aggregation.** The framework produces nine ordinal scores plus reliability modifiers without specifying how they compose into an overall assessment. This is a deliberate scope limitation: defining interaction functions requires a calibration dataset larger than currently available, and premature formalisation risks embedding assumptions that do not generalize. For production use, we expect expert judgment to integrate dimensional scores, paralleling early-stage TradFi rating framework development.

**Threats to Validity.** Dataset selection bias and size: the 12-incident illustration set is small relative to the 181 systematized DeFi attacks catalogued by Zhou et al. [22]. The set was assembled around incidents whose root cause exposes a structural feature relevant to dimensional design, which means the set naturally includes incidents that motivated the novel dimensions. Systematic mapping onto the Zhou et al. dataset is the necessary corrective. Back-fitting risk: the framework was developed by, among other inputs, studying these incidents. Coverage is therefore not surprising. We have used the term illustration rather than backtest or validation throughout §5. Genuine prospective validation requires application to incidents occurring after publication. Ontology-implementation gap: the ontological design is a design-principle description, not a published artifact. Independent reproduction at the conceptual level is intended; artifact-level reproduction is not. Single-jurisdiction framing: the regulatory analysis in Dimension 5 is dominated by U.S., EU, and Swiss frameworks. Substantial DeFi activity occurs in jurisdictions whose frameworks are not analyzed here.

**Independence and commercial relationships.** P3 (structural independence) is a governance commitment, not a mathematical property. The framework's architectural mechanisms — published framework, explainability chains, ordinal rubrics — create transparency that makes violations of independence detectable. They do not make violations impossible.

**Scope.** The framework assesses protocol-level risk. Position-level risk, portfolio-level risk, and market-level risk require additional analytical layers beyond this paper's scope.

# 8. Conclusion

This paper proposes a nine-dimension risk assessment framework for institutional DeFi, extending the 2022 taxonomy of Moody's Analytics and Gauntlet [4] with three novel dimensions: composability risk (D7), comprehension debt (D8), and temporal risk dynamics (D9). Constructive evidence from the 2024–2026 incident record demonstrates that each novel dimension captures risk information not derivable from the original six dimensions, and that the two highest-systemic-impact events in the illustrative dataset cannot be characterized without invoking the novel surfaces. We have framed the contribution explicitly as a proposal grounded in, but not validated by, the incident record: §5 is illustration, not backtest.

The framework makes three contributions. First, it provides the first explicit dimensional-independence target for DeFi risk assessment, with constructive evidence through observed incidents and orthogonality designated as a goal for empirical validation as the assessment dataset grows. Second, it introduces a transparency modifier grounded in traditional credit-rating architecture, operationalized through the Venus Principle. Third, it extends the class of comprehension debt — recently formalized in software engineering [24, 25–30] and empirically grounded in smart-contract practitioner studies [29, 33] — to the DeFi protocol context.

The framework rests on an extensible ontological design, described in §6 in design-principle form rather than published as an artifact. The design is part of the authors' ongoing protocol intelligence infrastructure, which is expected to evolve.

Five directions for future work follow directly. First, prospective validation: applying the framework to DeFi-relevant security incidents occurring after publication, recording dimensional assessments before incident outcomes are known, and reporting detection rates and false-positive rates. Second, systematic mapping onto the 181-incident SoK dataset of Zhou et al. [22], reporting on dimensional coverage across the broader incident population. Third, empirical validation of dimensional independence through factor-analytic methods as the population of assessed protocols grows. Fourth, operationalization of Dimension 8, developing protocol-level metrics that measurably separate inherent analyzability from audit quality. Fifth, applying the framework to the specific intersection of stablecoin regulation and DeFi composability, where the gap between issuer-level supervision and protocol-level risk propagation creates a regulatory need not addressed by current frameworks. A peer-reviewed validation of this framework will additionally require dimensional-independence testing across a population of 50 or more independently assessed protocols, prospective validation against incidents occurring after publication, an artifact-level ontological specification, and a measurement framework for Dimension 8 that operationally separates protocol analyzability from audit quality. The present paper is intended as a foundation for that work.